# Characterizing the 3D evolution of two successive CMEs heading for Mercury

Yanjie Zhang[1,*], Qingmin Zhang[1,*], Huadong Chen[2], Zhentong Li[1], Dong Li[1] and Haisheng Ji[1]

[1] Purple Mountain Observatory, Chinese Academy of Sciences, Nanjing 210023, China; *zhangyj@pmo.ac.cn*

[2] State Key Laboratory of Solar Activity and Space Weather, National Astronomical Observatories, Chinese Academy of Sciences, Beijing 100101, China

**Abstract** We studied two successive coronal mass ejections (CMEs) that erupted from the same active region (AR 12994) on 2022 April 15 and propagated toward Mercury. Using multi-view observations, we applied the revised cone model to determine the three-dimensional geometry and the early kinematics of the two CMEs. Our best-fit parameters indicate large angular extents of $84°$ and $86°$ and propagation directions of $-119.0°$ and $-110.4°$ (measured from the Sun–Earth line) for CME1 and CME2, respectively, while that of Mercury is $-120.1°$. The derived axis inclinations are $28°$ for CME1 and $21°$ for CME2, consistent with the orientation of the erupting flux ropes in the source region. Height–time analysis indicates approximately uniform motion speeds of 636 km s$^{-1}$ for CME1 and 696 km s$^{-1}$ for CME2, respectively. This paper provides valuable insights for predicting the impact of CMEs heading for Mercury as well as other solar planets in the future.

## 1 INTRODUCTION

Coronal mass ejections (CMEs) are large expulsions of plasma and magnetic fields from the solar atmosphere into interplanetary space (Forbes 2000; Webb & Howard 2012; Al-Haddad & Lugaz 2025). They carry large-scale magnetic flux and helicity, which are generally believed to originate from the destabilization and subsequent eruption of stressed or twisted coronal magnetic fields—often in the form of magnetic flux ropes—through magnetic reconnection or loss of equilibrium (Chen 2011; Priest et al. 2016; Warmuth & Mann 2020). A typical CME is characterized by a well-defined three-part structure comprising a bright core enclosed by a dark cavity and bounded by a bright leading edge (Illing & Hundhausen 1985; Song et al. 2019). CMEs with more complex morphologies have also been reported, such as two-component CMEs, comprising a narrow jet-like feature and a broader bubble-like structure produced by a single eruption (Shen et al. 2012; Duan et al. 2019, 2024). In recent years, CMEs have received growing attention owing to their profound influence on space weather near Earth, as well as at other planets throughout the heliosphere (Gold 1962; Plainaki et al. 2016; Temmer 2021; Palmerio et al. 2022; Soni et al. 2024).

---

*Corresponding Authors, these authors contributed equally to this work.

Understanding the three-dimensional (3D) morphology and propagation of CMEs is crucial for predicting their geo- or planet-effectiveness, while multi-view observational data must be utilized due to projection effects. A variety of geometric models have been developed over the past decades. Early studies introduced several versions of the cone model, which assume that CMEs propagate radially with constant angular width and velocity during their early expansion phase (Howard et al. 1982; Zhao et al. 2002; Michałek et al. 2003; Xie et al. 2004; Xue et al. 2005). These models were widely applied to determine the angular width, space speed, and arrival time of halo CMEs. With increasing evidence that many CMEs are driven by the eruption of magnetic flux ropes (Cheng et al. 2013; Yan et al. 2018), Thernisien et al. (2006) developed the Graduated Cylindrical Shell (GCS) model to reproduce the 3D morphology of flux-rope CMEs, providing a geometric framework that connects coronagraph observations with the underlying magnetic flux rope structure (Patsourakos et al. 2010; Thernisien 2011; O'Kane et al. 2021; Teng et al. 2024; Hu et al. 2025). Subsequent developments proposed more sophisticated approaches, which are designed to capture more detailed components and to reproduce the global 3D morphology and evolution of CMEs (Kwon et al. 2014; Isavnin 2016; Möstl et al. 2018).

To better constrain the 3D geometry and propagation characteristics of non-radial CMEs, Zhang (2021) extended the traditional cone model with two angles that describe the inclination of the CME axis relative to the local radial and meridional directions (see Section 2 for details). This revised formulation allows the cone apex to be located at the CME source region rather than at the Sun center, thereby accommodating lateral deflection and non-radial eruption trajectories. It has been successfully applied to wide-angle events, including halo and limb CMEs, enabling the determination of propagation direction, velocity, angular width and other key parameters (Zhang 2022; Zhang et al. 2024; Dai et al. 2023; Li et al. 2025).

On 2022 April 15, two successive CMEs erupted from the same active region (AR 12994) within ∼ 11 hours. The events have been investigated by Chen et al. (2024), which focused on the low-coronal evolution of the two successive eruptions, whereas the present study examines the early three-dimensional geometry and kinematics of the associated CMEs. Owing to the fortuitous separations between Earth, STEREO-A (STA), and Solar Orbiter (SolO), the events were observed from three well-separated vantage points. Based on these simultaneous observations, the 3D kinematics of both CMEs were reconstructed using the revised cone model. Remarkably, both CMEs were found to propagate in directions close to Mercury. This work may provide some basis for future investigations into the impact of CMEs on planetary space weather.

The paper is organized as follows. Section 2 describes the data and the revised cone model. Section 3 details the geometric reconstruction and the analysis of the CMEs. Discussion and summary are given in Section 4.

## 2 DATA AND METHOD

Multi-instrument observations are used to investigate the propagation of the CMEs from the low corona to interplanetary space. The Large Angle and Spectrometric Coronagraph (LASCO; Brueckner et al. 1995) on board the SOHO spacecraft provides white-light observations of the corona. In this work, we employed data from the C2 coronagraph, which observes the corona in the range of 1.5–6.0 $R_\odot$ with a typical cadence of about 12 min.

The Sun Earth Connection Coronal and Heliospheric Investigation (SECCHI; Howard et al. 2008) suite aboard the STEREO-Ahead (STA) spacecraft includes several complementary instruments. The Extreme-Ultraviolet Imager (EUVI; Wuelser et al. 2004) observes the full solar disk in four EUV passbands (171, 195, 284, and 304 Å), covering plasma temperatures from $1 \times 10^5$ to $2 \times 10^7$ K, with a field of view (FOV) extending to $\sim 1.7\,R_\odot$ and a cadence of 2.5–10 min. The outer coronagraph COR2 extends the FOV to $15\,R_\odot$ with a cadence of about 15 min.

Data from the Solar Orbiter mission (SolO; Müller et al. 2020) are also used. The Extreme Ultraviolet Imager (EUI; Rochus et al. 2020) includes a Full Sun Imager (FSI) and two High Resolution Imagers (HRIs) operating at Lyman-$\alpha$ and EUV (174 Å, 304 Å). The FSI has a $3.8° \times 3.8°$ FOV, while the HRI provides high spatial resolution imaging of small-scale coronal structures. The Spectrometer Telescope for Imaging X-rays (STIX; Krucker et al. 2020) observes solar X-rays in the 4–150 keV range, providing imaging spectroscopy of flare-accelerated electrons and the associated energy release. Together, EUI and STIX offer complementary EUV and X-ray diagnostics of solar eruptive events from a unique vantage point away from the Sun–Earth line.

The Atmospheric Imaging Assembly (AIA; Lemen et al. 2012) on board the Solar Dynamics Observatory (SDO; Pesnell et al. 2012) provides full-disk EUV and UV imaging with a spatial resolution of $1.2''$ and a cadence of 12 s. AIA observes in seven EUV passbands ranging from 94 to 335 Å, covering multiple temperature regimes of the solar atmosphere.

To investigate the three-dimensional geometry and kinematics of the CMEs resulting from non-radial eruptions, we adopted the revised cone model (Zhang 2021, 2022; Zhang et al. 2024). In contrast to the traditional cone models that assume a radial symmetry and place the cone apex at the Sun center, the revised version locates the apex at the eruption source region on the solar surface and allows the cone axis to deviate from the local vertical by an inclination angle $\theta_1$ and an azimuthal deviation $\phi_1$. The cone is characterized by four geometric parameters: the axial length $r$, the angular width $\omega$, the inclination $\theta_1$, and the deviation $\phi_1$. These parameters fully determine the CME's spatial orientation and size in three dimensions.

The total length of the leading edge is then

$$l = r \left( \tan \frac{\omega}{2} + \sec \frac{\omega}{2} \right), \tag{1}$$

To compare the model with observations, the cone geometry is forward-projected into each observer's plane of the sky using spacecraft-specific geometry matrices. This geometric forward modeling provides the instantaneous three-dimensional morphology and propagation direction of CMEs during their early evolution.

## 3 RESULTS

Figure 1 illustrates the relative positions of the Sun, Earth, Mercury, STA and SolO on 2022 April 15 at 01:00 UT, in which two successive CMEs (CME1 and CME2) were investigated using multi-viewpoint observations from LASCO, STA, and SolO within the framework of the revised cone model (Zhang 2021, 2022; Zhang et al. 2024). STA and SolO have separation angles of $-32.1°$ and $146.0°$ with the Sun-Earth line, respectively. The separation angle of Mercury with respect to Earth was $-120.1°$. The two black arrows pointing toward Mercury indicate the directions of CME1 and CME2 derived from the revised cone model, respectively. A detailed explanation is provided below.

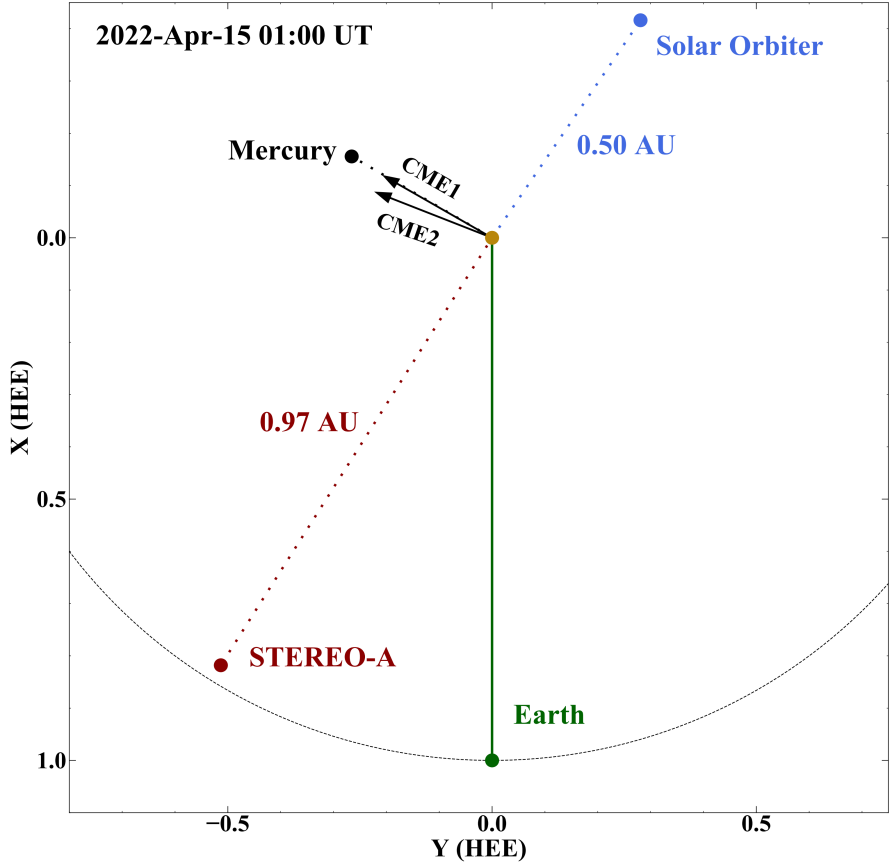

Fig. 1: Schematic geometry on 2022 April 15 at 01:00 UT. The positions of the Sun (center), Earth, Mercury, STA, and SolO are projected onto the ecliptic plane. The Sun–Earth line is vertical. Black arrows indicate the fitted propagation directions of CME1 and CME2, respectively.

Figure 2 presents high-resolution EUV observations of AR 12994 in the low corona (SDO/AIA 131 Å and 171 Å). Panels (a1)–(a3) document the early impulsive signatures associated with the first flare (Flare1), where compact, hot loops or blobs rapidly expands and detaches, while panels (b1)–(b3) show the overlying loops of the second flare (Flare2) exhibiting both expansion and rapid shrinkage sequences. The morphological patterns—rapid loop lifting (E-loops in panel (b1)) followed by loop contraction (I-loops1 and I-loops2 in panel (b1)) during successive energy releases—are detailed in the case study by Chen et al. (2024). The AIA observations directly link the low-coronal energy release to the mass evacuation (the dimming shown in the EUI observations presented below) and the early-stage geometry of the CME as represented by the cone model: CME1 is more inclined than CME2 along the $Z_l$ aixs in the POS ($\theta_1$ equals $28°$ for CME1 and 21° for CME2, as detailed in the following text), coinciding with a more inclined erupting loop of Flare1 than that of Flare2, as shown by the arrows in Figure 2. This correspondence supports a picture in which the flare-mediated restructuring of the low coronal magnetic field plays a controlling role in shaping the initial CME geometry and its early kinematics.

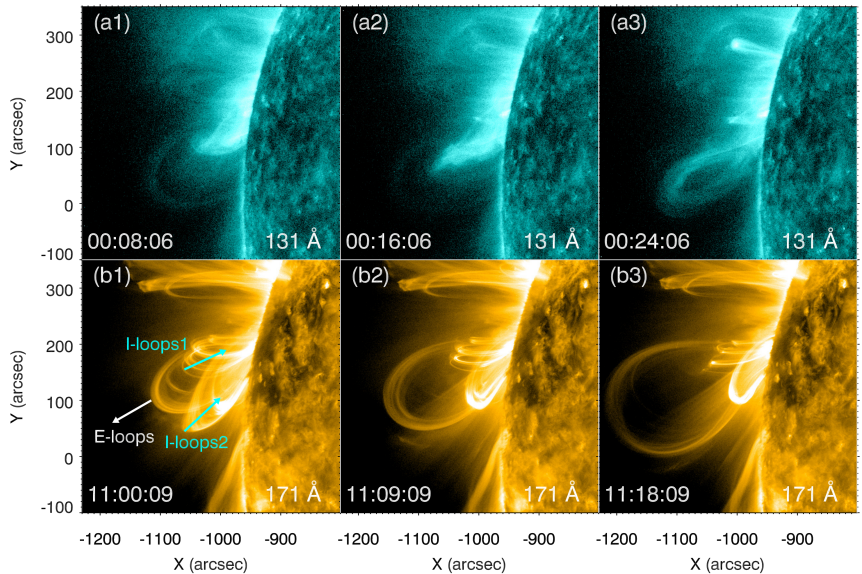

Fig. 2: AIA observations of AR 12994. (a1)–(a3) AIA 131 Å images from 00:08 to 00:24 UT showing Flare1 loops or blobs rapidly expanding during Flare1. (b1)–(b3) AIA 171 Å images from 11:00 to 11:18 UT showing overlying loops rising (white arrow) and contracting (cyan arrows) during Flare2.

Figure 3 presents the source regions of the two eruptions observed by the EUVI at 195 Å and 304 Å. Panels (a1)–(b1) show Flare1 at 01:00 UT, while panels (a2)–(b2) depict Flare2 at 11:25 UT. In the 195 Å images, the coronal arcades brightened and expanded rapidly, while the 304 Å images reveal the eruption of cool prominence material overlying the flare site. The similar morphology and location of the two flares indicate that they occurred within a homogeneous magnetic system. As indicated in Figure 2 and Figure 3, these two CMEs originate from the non-radial eruption of coronal loop or hot channel, with their footpoints anchored in AR 12994. Consequently, they are quite suitable for investigation using the revised cone model. However, it should be noted that the model is also applicable to radial CMEs, in which case the inclination angles $\theta_1$ and $\phi_1$ reduce to zero.

The temporal evolution of the two CMEs and their associated flares observed on 2022 April 15 by SolO/STIX are presented by Figure 4. Panels (a) and (b) show the X-ray flux profiles in the energy ranges of 4–10 keV (blue), 10–15 keV (orange), and 15–25 keV (green), together with the height–time measurements of the corresponding CMEs derived from the revised cone model (magenta symbols). Flare1 peaked at ∼01:00 UT, followed by CME1, whose leading-edge height ($h_{\rm LE}$) increased from 2.35 $R_\odot$ to 13.86 $R_\odot$ during 01:24–04:54 UT. Flare2 peaked at ∼11:23 UT, accompanied by CME2 observed in the FOV covering 2.53 $R_\odot$ to 14.19 $R_\odot$ during 12:04 UT -15:24 UT. A linear fit to the height–time data yields approximately uniform motion for the two CMEs, with mean velocities of 636 km s$^{-1}$ for CME1 and 696 km s$^{-1}$ for

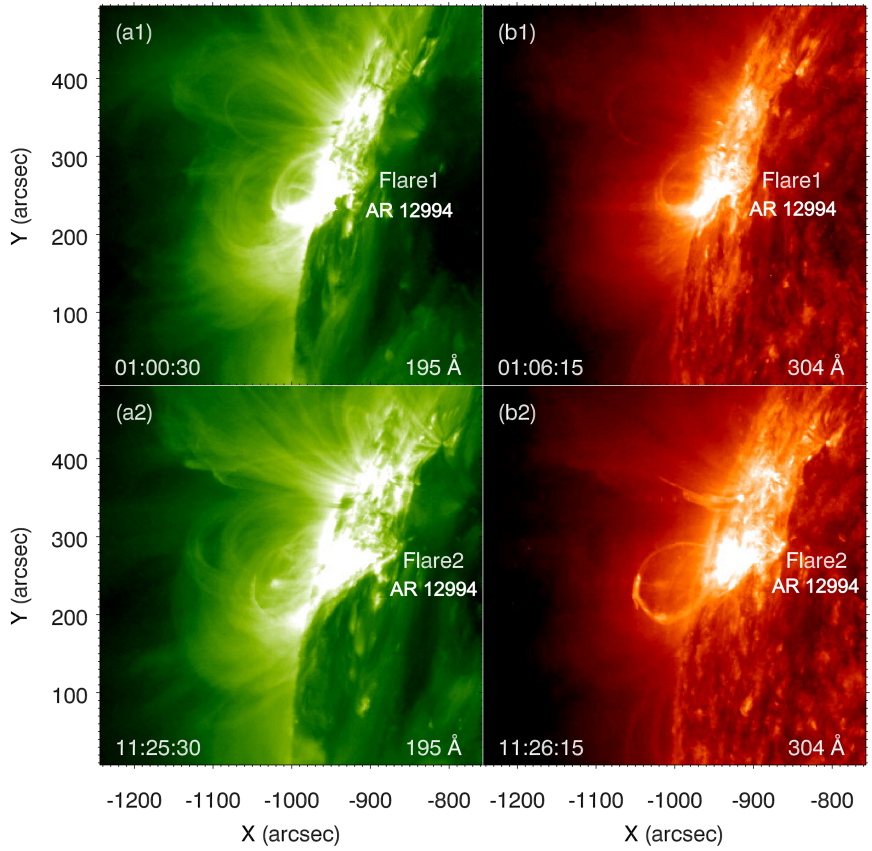

Fig. 3: STA/EUVI observations of AR 12994. (a1)–(b1): Flare1 at 01:00 UT in 195 Å and at 01:06 UT in 304 Å; (a2)–(b2): Flare2 at 11:25 UT in 195 Å and at 11:26 UT in 304 Å.

Table 1: Parameters of the Cones at Different Times of CME1.

| Time (UT) | 01:24 | 01:36 | 01:39 | 01:48 | 01:54 | 02:00 | 02:09 | 02:12 | 02:24 | 02:39 |
|---|---|---|---|---|---|---|---|---|---|---|
| $r\ (R_\odot)$ | 2.35 | 2.82 | 2.94 | 3.29 | 3.64 | 3.81 | 4.35 | 4.46 | 5.17 | 5.99 |
| Time (UT) | 02:54 | 03:09 | 03:24 | 03:39 | 03:54 | 04:09 | 04:24 | 04:39 | 04:54 | |
| $r\ (R_\odot)$ | 6.81 | 7.64 | 8.41 | 9.05 | 9.99 | 10.92 | 11.98 | 12.69 | 13.86 | |

**Notes.** The values of $\phi_2 = -120°$, $\theta_2 = 77°$, $\phi_1 = 0°$, $\theta_1 = 28°$, and $\omega = 84°$ keep constant during the evolution.

CME2, respectively. The temporal correspondence between the flare peaks and the CME onsets is additional evidence for a close physical connection between the impulsive energy release in the low corona and the outward expansion of the CME front (Chen et al. 2024). The parameters derived from the revised cone model are listed in Table 1 and Table 2.

Figure 5 shows the multi-viewpoint observations of CME1 obtained nearly simultaneously by LASCO-C2, STA/COR2, and SolO/EUI. Panels (a1)–(a2), (b1)–(b2), and (c1)–(c2) correspond to the views from the Earth, STA, and SolO, respectively. The upper panels present the original running-difference images,

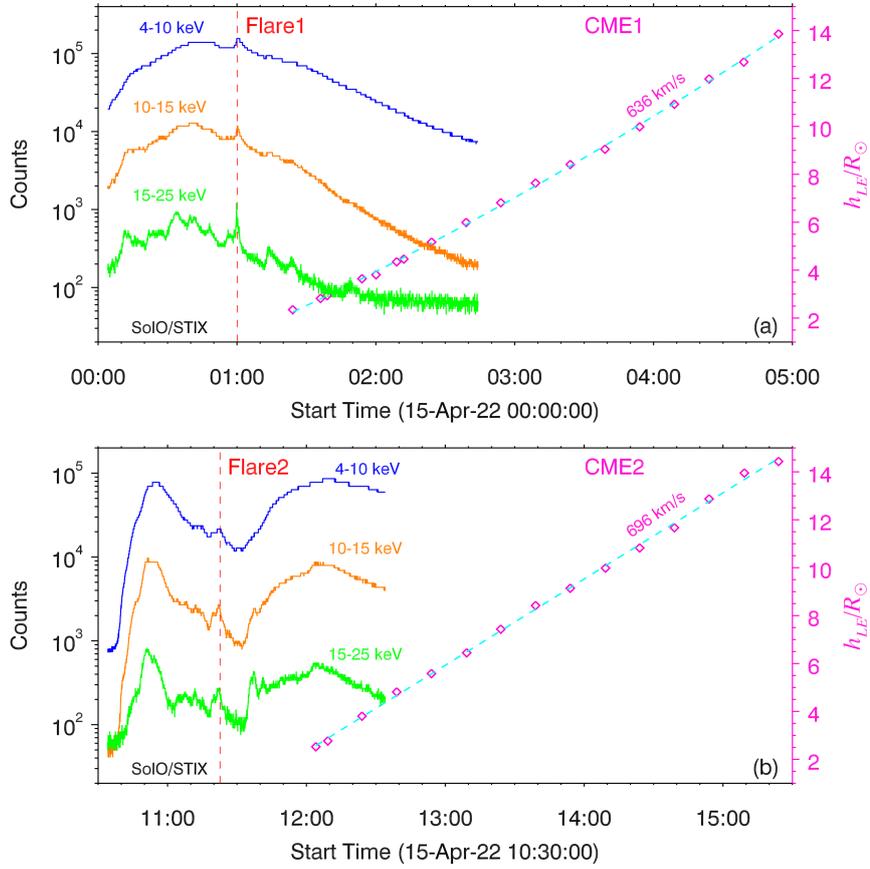

Fig. 4: Temporal evolution of the flare X-ray emission and CME heights on 2022 April 15. (a) Flare1 (blue/orange/green: STIX 4–10/10–15/15–25 keV) and the leading-edge height of CME1 (magenta points). (b) The same as in panel (a) for Flare2 and CME2. Linear fits (dashed lines) yield $v \simeq 636$ m s$^{-1}$ for CME1 and $v \simeq 696$ km s$^{-1}$ for CME2, respectively.

Table 2: Parameters of the Cones at Different Times of CME2.

| Time (UT) | 12:04 | 12:09 | 12:24 | 12:39 | 12:54 | 13:09 | 13:24 | 13:39 | 13:54 | 14:09 |
|---|---|---|---|---|---|---|---|---|---|---|
| $r$ ($R_\odot$) | 2.53 | 2.77 | 3.80 | 4.81 | 5.58 | 6.45 | 7.43 | 8.42 | 9.14 | 9.98 |
| Time (UT) | 14:24 | 14:39 | 14:54 | 15:09 | 15:24 | | | | | |
| $r$ ($R_\odot$) | 10.83 | 11.67 | 12.51 | 13.35 | 14.19 | | | | | |

**Notes.** The values of $\phi_2 = -114.2°$, $\theta_2 = 77°$, $\phi_1 = 10°$, $\theta_1 = 21°$, and $\omega = 86°$ keep constant during the evolution.

while the lower panels display the same frames with the fitted revised cone model projected onto the POS. At 01:24 UT, CME1 appeared as a bright front expanding southeast in both the LASCO-C2 and COR2 images. In the SolO/EUI image taken at 01:10:45 UT, a pronounced coronal dimming is evident at the CME source region (indicated by the white arrow in panel (c1)), suggesting the evacuation of coronal plasma during the early eruption stage. This dimming region corresponds spatially to the flare site analyzed in the study of simultaneous loop eruption reported by Chen et al. (2024), confirming that CME1 originated from the same active region involved in the flare–CME coupling process. The projection of the revised cone model (cyan mesh) reproduces the observed CME envelope well in all three viewpoints, indicating the reliability of the derived geometrical parameters. The consistent morphology among LASCO, STA, and SolO implies that the CME1 maintains a coherent three-dimensional structure during its early expansion.

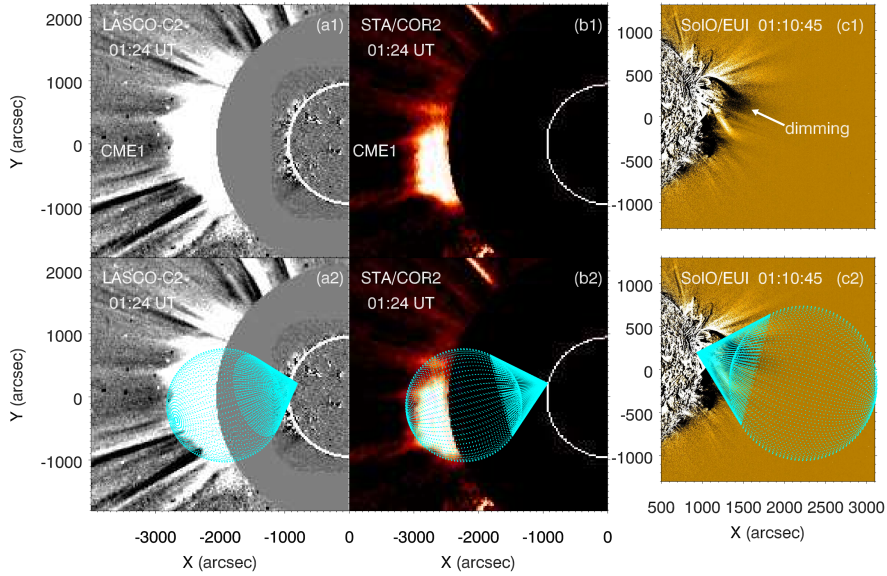

Fig. 5: Multi-perspective observations of CME1 at about 01:24 UT on 2022 April 15. Panels (a1)–(c1) are running-difference images from LASCO-C2 (Earth view), STA/COR2, and SolO/EUI, respectively. Panels (a2)–(c2) show the same frames with the fitted revised cone model (cyan mesh) superimposed. The white arrow in panel (c1) points to coronal dimming at the source region.

Figure 6 presents another set of simultaneous observations of CME1 at 02:24 UT from LASCO-C2 (left panels) and STA/COR2 (right panels). The bright leading edge, the dark cavity, and the bright core of the CME are clearly visible, forming a well-defined three-part morphology. Panels (a2) and (b2) show the fitted cone model (magenta mesh) superimposed on the observed CME fronts. The model reproduces both the projected size and the curvature of CME1. Compared with the earlier time shown in Figure 5, CME1 has undergone significant radial expansion and brightening, consistent with its acceleration phase inferred from the height–time analysis (see Figure 4). The close agreement between the model and the observed morphology further validates the reliability of the revised cone geometry in representing the three-dimensional evolution of CME.

Figure 7 displays near-simultaneous observations of CME2 at 12:54 UT as seen by LASCO-C2 (left column) and STA/COR2 (right column). Panels (a1)–(b1) present the running-difference images showing the bright leading edge, while panels (a2)–(b2) overlay the three-dimensional revised-cone mesh (magenta mesh) on the same frames. The cone projection reproduces the observed outline of CME2 front and its lateral extent reasonably well in both perspectives. The bright, nearly circular leading front in LASCO-C2 is consistent with the modeled spherical-cap geometry, whereas the STA/COR2 view reveals a more lobate and structured interior, indicating density inhomogeneities along the flank. The asymmetric brightness suggests a combination of line-of-sight projection and intrinsic non-uniform mass distribution within the erupting flux system. Overall, the agreement between the cone model and multi-viewpoint images validates the derived geometrical parameters for CME2 and supports the interpretation that the observed morphology results primarily from projection of a coherent, expanding bubble with localized density enhancements.

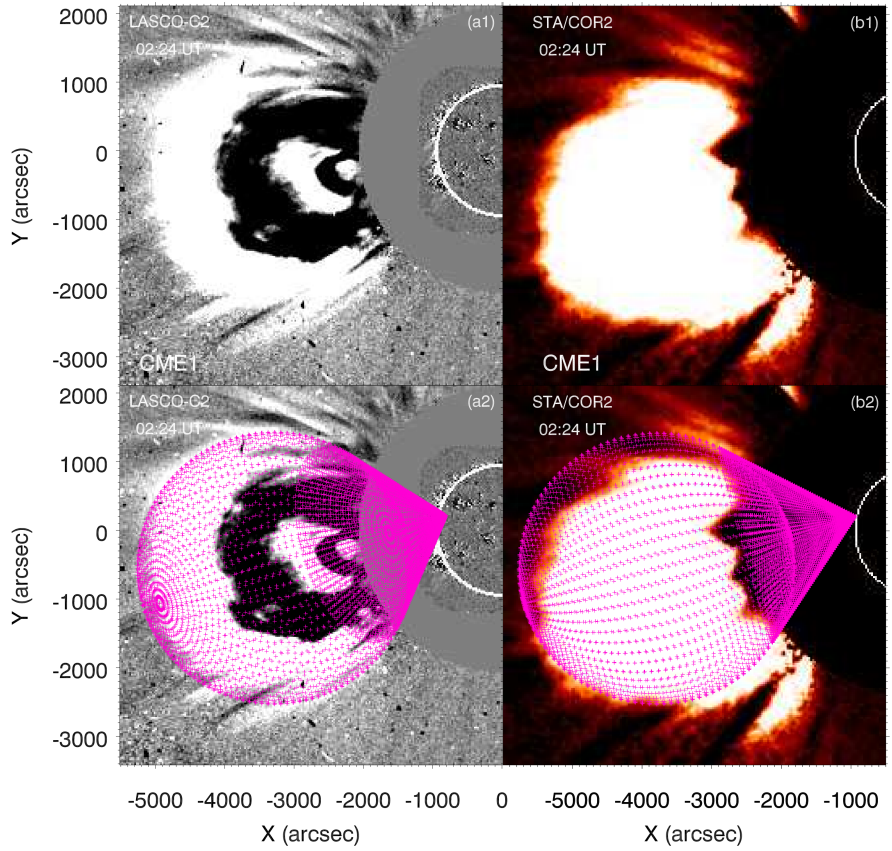

Fig. 6: CME1 at 02:24 UT. Left: LASCO-C2 images; right: STA/COR2 images. Top row: running-difference images. Bottom row: the same images with the fitted cone model (magenta mesh) overlaid.

The temporal evolution of both CMEs from the STA/COR2 viewpoint is summarized by Figure 8, with cyan meshes representing the revised-cone projection at selected times. Panels (a)–(d) follow CME1 at 01:54, 02:54, 03:54 and 04:54 UT, while Panels (e)–(h) present CME2 at 12:09, 13:09, 14:09 and 15:09 UT, both showing a clear progression from a compact, bright front to a more extended, diffuse envelope (The mean pixel intensities of the two CMEs decreased from 202 to 161 for CME1, and from 205 to 167 for CME2). During the observation interval, the fitted axial length $r$ of CME1 increased from approximately $3.64\,R_\odot$ to $13.86\,R_\odot$, while that of CME2 grew from about $2.77\,R_\odot$ to $13.35\,R_\odot$. By contrast, the angular widths remained essentially unchanged, with $\omega = 84°$ for CME1 and $\omega = 86°$ for CME2.

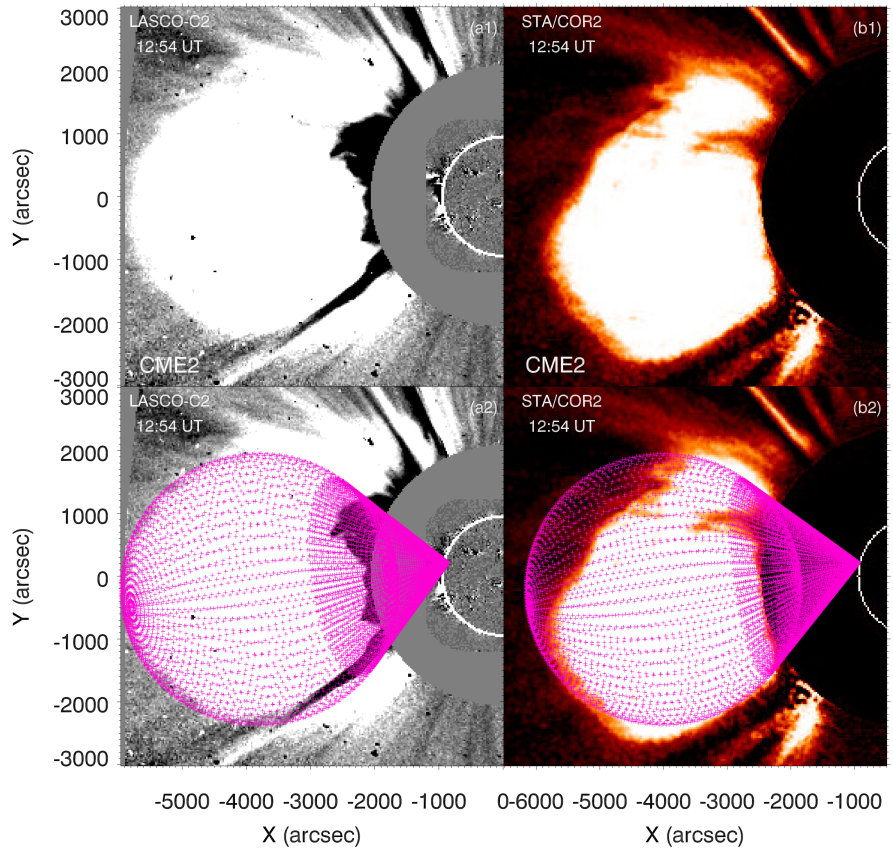

Fig. 7: CME2 at 12:54 UT. Left: LASCO-C2 images; right: STA/COR2 images. Top row: running-difference images. Bottom row: the same images with the fitted cone model (magenta mesh) overlaid.

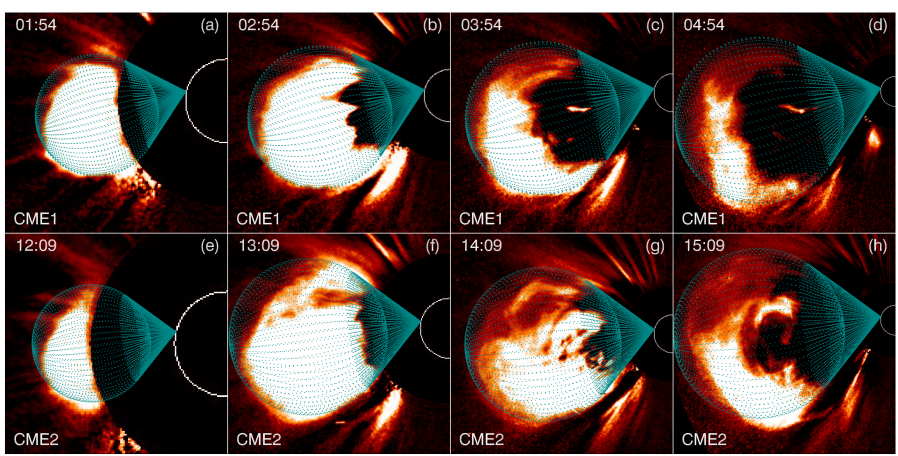

Fig. 8: Time sequence of the CMEs in STA/COR2 with the cone-model fits (cyan meshes). (a)–(d) CME1 at 01:54, 02:54, 03:54, 04:54 UT; (e)–(h) CME2 at 12:09, 13:09, 14:09, 15:09 UT. The cone model captures the large-scale expansion of each CME front.

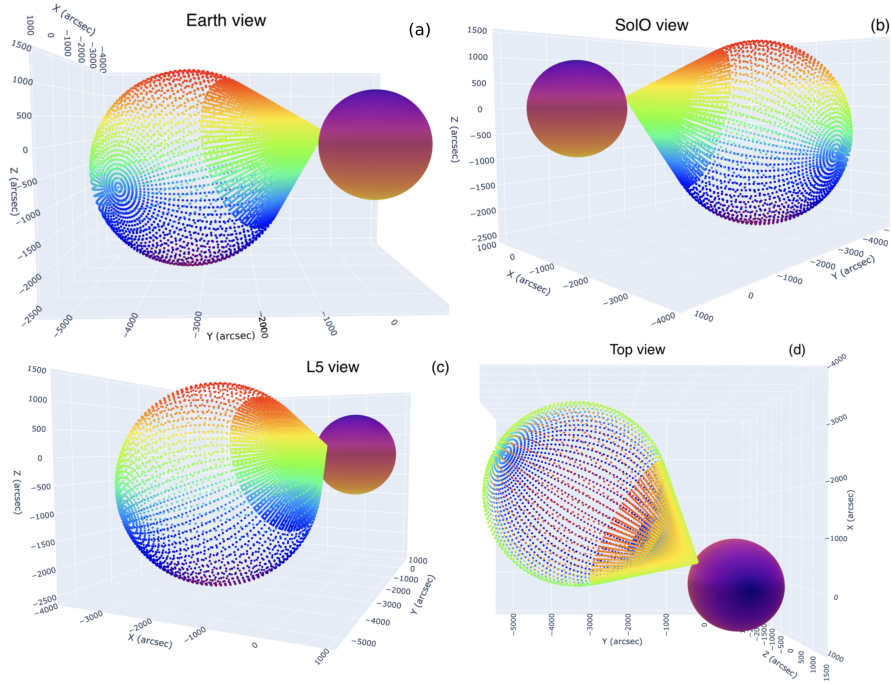

Fig. 9: Three-dimensional visualization of the reconstructed CME1 at 02:24 UT, shown from four different viewpoints: (a) Earth, (b) SolO, (c) the Sun–Earth L5 Lagrange point, and (d) a polar-orbit vantage.

The cone model captures the overall radial expansion and the changing apparent curvature of the front, while deviations between the model and observation—most notably the evolution of thread-like outward brightness features along the flank—point to non-axisymmetric density redistribution during the propagation process. From the fitted results, the propagation directions of the two CMEs form angles of $-119.0°$ and $-110.4°$ with the Sun–Earth line (see Figure 1), respectively. It is worth noting that the inclination angles among the propagation directions of the two CMEs, the orbital plane of Mercury and the ecliptic plane of Earth are sufficiently small to be neglected (approximately $15.0°$ for CME1, $7.7°$ for CME2, and $7.0°$ for the orbital plane of Mercury with respect to the ecliptic plane of Earth).

## 4 DISCUSSION AND SUMMARY

The source regions of the two consecutive CMEs have been investigated in detail by Chen et al. (2024). In their study, the flare loops from the first eruption were swept up into the second eruption: the central (E-loops) part of the loops became part of the new erupting flux rope, while the side loops (I-loops1, I-loops2) initially rose and then contracted by tens of megameters. Such loop contraction at speeds of tens to ∼100 km s$^{-1}$ is a signature of large-scale magnetic energy converted and released (Fletcher et al. 2013; Priest 2014). Compared with the extreme non-radial events reported by Zhang (2021), the two CMEs studied here exhibit moderate inclination angles and relatively stable deviations, more comparable to the later-stage configurations reported in Zhang (2022) and Li et al. (2025). Furthermore, the axis inclinations of the CME1 ($\theta_1 = 28°$) and the CME2 ($\theta_1 = 21°$) exhibit a correlation with the erupting flux rope orientations, in that the more highly tilted rope associated with Flare1 corresponds to the greater axis inclination of the CME1.

The second eruptive flux rope were flanked on both sides by other magnetic loop structures (I-loops), which may significantly confined or modulated the eruption direction during the eruptive process. This supports the idea that the flare reconnection geometry imprints on the launch direction of CME. In other words, as the low-coronal field restructures during the flare, it may bias the subsequent CME trajectory.

On the other hand, while most CME studies have focused on Earth-directed events, this work reinforces the challenge of space weather extending to other planets in the solar system, not only Earth. In this paper, both CMEs were heading for Mercury, despite appearing as limb events from Earth. Since CME acceleration occurs primarily within the low coronal region (Bein et al. 2011), the two CMEs are assumed to propagate at constant speeds after our observational period (with average velocities of 636 km s$^{-1}$ for CME1 and 696 km s$^{-1}$ for CME2). Accordingly, their predicted arrival times at Mercury are about 06:00 UT and 14:00 UT on 16 April, respectively. Given the weak magnetosphere of Mercury, even moderate CMEs can have extreme effects on it (Slavin et al. 2014; Exner et al. 2018; Winslow et al. 2020; Sun et al. 2020). It is pointed out that in situ measurements at the distance of Mercury can differ drastically even over small separations (Palmerio et al. 2024). Furthermore, this work also provides a foundation for the achievement of the scientific objectives of China's upcoming Xihe-2 (which will observe from the Sun-Earth L5 Lagrange point, with its viewing geometry shown in Figure 9(c)) and Kuafu-2 (which will operate in a solar polar orbit, with its viewing geometry shown in Figure 9(d)) satellites.

In this paper, we have carried out a comprehensive 3D reconstruction of two successive CMEs on 2022 April 15, using the revised cone model with multi-spacecraft observations. The main results are as follows:

Both CMEs originated from the same limb active region (AR 12994) and were closely associated with two successive flares (Flare1 peaked at about 01:00 UT, Flare2 at about 11:23 UT). The flare loop formed during Flare1 was involved in Flare2. The middle part erupted outward in conjunction with the underlying flux rope, while the lateral parts first rose and then contracted, as reported by Chen et al. (2024). The derived axis inclination of CME1 ($\theta_1 = 28°$) is steeper than that of CME2 ($\theta_1 = 21°$), correlating with the orientation of the erupting flux ropes.

CME1 and CME2 had large angular widths (84° for CME1, 86° for CME2) and propagated toward Mercury ($-119.0°$ for CME1 and $-110.4°$ for CME2 from the Sun–Earth line, while $-120.1°$ for Mercury). Both CMEs propagated at an almost constant speed of 636 km s$^{-1}$ for CME1 and for 696 km s$^{-1}$ for CME2.

In conclusion, by combining high-resolution imaging, multi-vantage point coronagraph data, and the revised cone model, we have characterized the 3D evolution of two near-Mercury CMEs. Our reconstructions thus provide valuable input for predicting their impact on Mercury as well as other solar planets in the future.

**Acknowledgements** The authors are grateful to the colleagues in Purple Mountain Observatory and National Astronomical Observatories for their constructive suggestions and comments. SOHO is a cooperative ESA–NASA spacecraft. SolO is a mission of ESA with contributions from NASA and other international partners. SDO is a mission of NASA's Living With a Star Program. STEREO is a mission of NASA's Solar Terrestrial Probes program. This work was funded by the National Natural Science Foundation of China (NSFC) under No.12403068, 12573057; Natural Science Foundation of Jiangsu

Province (BK20241707). H.C. is supported by the National Key R&D Program of China 2021YFA1600502, the Strategic Priority Research Program of the Chinese Academy of Sciences (grant No. XDB0560000), the National Natural Science Foundations of China (12533010), and the Specialized Research Fund for State Key Laboratories of Solar Activity and Space Weather.